\documentclass[twocolumn,floatfix]{revtex4}
\usepackage{graphicx}
\usepackage{amsmath}
\usepackage{amssymb}
\usepackage{bm}

\begin{document}

\title{Proposal for the detection and braiding of Majorana fermions in a quantum spin Hall insulator}
\author{Shuo Mi}
\affiliation{Instituut-Lorentz, Universiteit Leiden, P.O. Box 9506, 2300 RA Leiden, The Netherlands}
\author{D. I. Pikulin}
\affiliation{Instituut-Lorentz, Universiteit Leiden, P.O. Box 9506, 2300 RA Leiden, The Netherlands}
\author{M. Wimmer}
\affiliation{Instituut-Lorentz, Universiteit Leiden, P.O. Box 9506, 2300 RA Leiden, The Netherlands}
\author{C. W. J. Beenakker}
\affiliation{Instituut-Lorentz, Universiteit Leiden, P.O. Box 9506, 2300 RA Leiden, The Netherlands}

\date{April 2013}
\begin{abstract}
We show how a quantum dot with a ballistic single-channel point contact to a superconductor can be created by means of a gate electrode at the edge of a quantum spin Hall insulator (such as an InAs/GaSb quantum well). A weak perpendicular magnetic field traps a Majorana zero-mode, so that it can be observed in the gate-voltage-averaged differential conductance $\langle dI/dV\rangle$  as a $4e^{2}/h$ zero-bias peak above a $(\frac{2}{3}\pi^{2}-4)e^{2}/h$ background. The one-dimensional edge does not permit the braiding of pairs of Majorana fermions, but this obstacle can be overcome by coupling opposite edges at a constriction, allowing for a demonstration of non-Abelian statistics.
\end{abstract}
\maketitle

Topological insulators in proximity to a superconductor have been predicted \cite{Fu08} to support Majorana zero-modes: midgap states with identical creation and annihilation operators and non-Abelian braiding statistics \cite{Kit01,Rea00}, that are presently under intense scrutiny \cite{Ali12}. The conducting edge of a quantum spin Hall (QSH) insulator seems like an ideal system to search for these elusive particles in a transport experiment \cite{Fu09,Nil08}: Only a single mode propagates in each direction along the edge, unaffected by disorder since backscattering of these helical modes is forbidden by time-reversal symmetry \cite{Has10}. The QSH edge is thus immune for the multi-mode and disorder effects that complicate the Majorana-fermion interpretation of transport experiments in semiconductor nanowires \cite{Fra03,Sta03}.

Andreev reflection at a superconducting interface has been reported in an InAs/GaSb quantum well \cite{Kne12}, which is a QSH insulator because of a band inversion and the appearance of edge states connecting conduction and valence bands \cite{Liu08}. Similar experiments can be tried in HgTe/CdTe quantum wells, where the QSH effect was first discovered \cite{Ber06,Kon07}. We expect a Majorana fermion to be present in these systems, delocalized along the edge connecting a normal and superconducting contact, but without a distinctive resonance in the electrical conductance. Andreev reflection of a helical edge mode doubles the current at all energies inside the band gap, so each edge contributes $2e^2/h$ to the differential conductance irrespective of any midgap states.

\begin{figure}[tb]
\centerline{\includegraphics[width=0.8\linewidth]{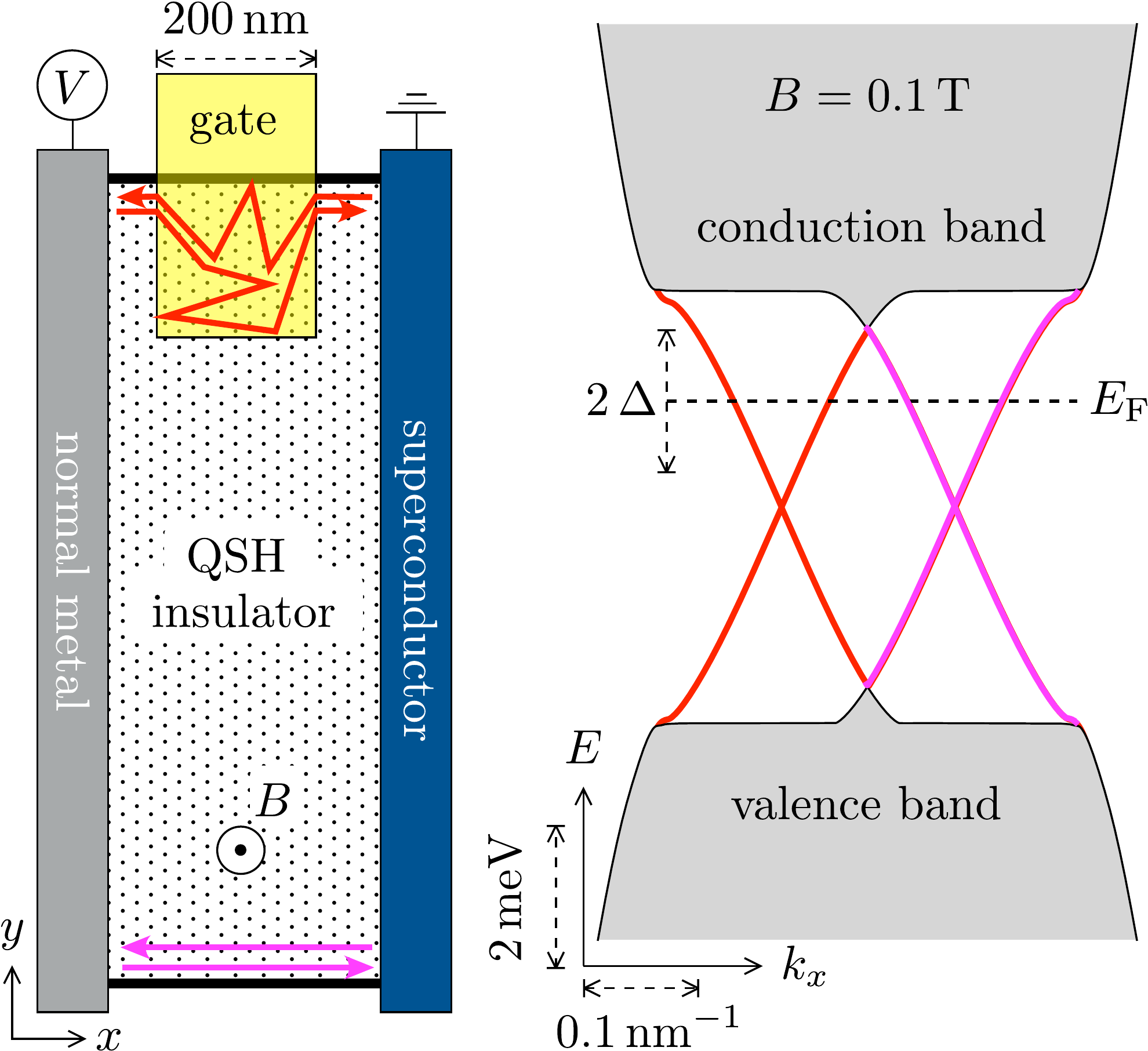}}
\caption{\textit{Left panel:} Andreev quantum dot, created by a gate electrode at the edge of a quantum spin Hall (QSH) insulator in a perpendicular magnetic field $B$. A current $I$ is passed between metallic and superconducting contacts, and the differential conductance $G=dI/dV$ is measured as a function of the bias voltage $V$ for different gate voltages. \textit{Right panel:} Band structure of an InAs/GaSb quantum well, for the parameters used in the computer simulations. The helical edge states appear inside the gap, connecting conduction and valence bands. Below the gate, the conduction band is pushed through the Fermi level at $E_{\rm F}$, to create a metallic puddle. Inside the superconducting contact, a gap $2\Delta$ opens around the Fermi level.
}
\label{fig_setup}
\end{figure}

Here we present a method to restore the sensitivity of the conductance to the zero-mode resonance, by trapping the Majorana fermion near the superconducting interface. Only a minor modification of the existing experimental setup \cite{Kne12} is needed, essentially only a gate electrode at one of the edges, to locally push the conduction band through the Fermi level. (See Fig.\ \ref{fig_setup}.) The area under the gate then forms a two-dimensional metallic region, connected to the superconductor by the helical edge mode. Backscattering at this Andreev quantum dot in a weak magnetic field (one flux quantum or less through the dot) provides for an electrostatically tunable confinement of Majorana fermions. We discuss the detection of Majoranas as a short-term application, and braiding as a longer term perspective.

There exists a variety of phase coherent backscattering mechanisms for helical edge modes \cite{Mac10,Tan11,Hat11,Il12,Del12,Tim12,Mae12}, based on different methods of time-reversal symmetry breaking to open a minigap in the edge state spectrum. A locally opened minigap forms a tunnel barrier for the edge modes and two tunnel barriers in series form a quantum dot at the QSH edge \cite{Tim12}. For a robust Majorana resonance it is advantageous to have a ballistic coupling rather than a tunnel coupling to the superconductor, so we form a quantum dot by placing two ballistic point contacts in series --- without opening an excitation gap at the Fermi level. 

The geometry, sketched in Fig.\ \ref{fig_setup}, can be seen as a gate-controlled realization of the puddles of metallic conduction that may occur naturally near the QSH edge \cite{Rot09,Kon12,Vay13}. An electron entering the metallic area under the gate from one side can be either transmitted to the other side or reflected back to the same side, with amplitudes contained in the $2\times 2$ unitary scattering matrix $S(\varepsilon)$, dependent on the energy $\varepsilon$ relative to the Fermi level. Time reversal symmetry requires an antisymmetric scattering matrix \cite{Bar08}, $S_{nm}=-S_{mn}$, so the reflection amplitudes on the diagonal are necessarily zero and the gate has no effect on the conductance. 

A perpendicular magnetic field $B$ effectively removes this constraint, once the flux through the gate is of the order of a flux quantum $h/e$. 
The electronic scattering matrix then has the four-parameter form
\begin{equation}
\begin{split}
&S=\begin{pmatrix}
r'&t'\\
t&r
\end{pmatrix}=e^{i\phi_{1}\sigma_{0}}e^{i\phi_{2}\sigma_{z}}e^{i\gamma\sigma_{y}}
e^{i\phi_{3}\sigma_{z}},\\
&\gamma\in[0,\pi/2),\;\;\phi_{n}\in[0,2\pi),\;\;n=1,2,3.
\end{split}\label{Sgeneric}
\end{equation}
We have introduced Pauli spin matrices $\sigma_{x},\sigma_{z}$, with $\sigma_{0}$ the $2\times 2$ unit matrix.

If the scattering in the quantum dot is chaotic, the matrix $S$ is uniformly distributed among all $2\times 2$ unitary matrices. The Haar measure on the unitary group gives the probability distribution
\begin{equation}
P(\gamma,\phi_{1},\phi_{2},\phi_{3})=(2\pi)^{-3}\sin 2\gamma,\label{Pphidef}
\end{equation}
representing the circular unitary ensemble (CUE) of random-matrix theory \cite{Bee97}. This produces a transmission probability $T=|t|^{2}=\sin^{2}\gamma$ that is \textit{uniformly} distributed between zero and one \cite{Bar94,Jal94}. Different realizations of the ensemble, with different $T\in[0,1]$, can be reached by varying the gate voltage, so that the quantum dot in a magnetic field functions as a tunable transmitter for the helical edge channels.

We now use this quantum dot as an energy-sensitive detector of the presence of a Majorana zero-mode at the interface with a superconductor. To explain how the energy sensitivity appears, we follow the usual procedure \cite{Bee97} of combining the electronic scattering matrix $S(\varepsilon)$, the hole scattering matrix $S^{\ast}(-\varepsilon)$, and the Andreev reflection matrix
\begin{equation}
r_{\rm A}=\alpha\tau_{y},\;\;\alpha=\sqrt{1-(\varepsilon/\Delta_{0})^{2}}+i\varepsilon/\Delta_{0}.\label{rAdef}
\end{equation}
The Pauli matrix $\tau_{y}$ acts on the electron-hole degree of freedom and $\Delta_{0}$ is the superconducting gap. An electron incident on the quantum dot along a helical edge state is reflected back as a hole with probability
\begin{equation}
R_{he}(\varepsilon)=\frac{T(\varepsilon)T(-\varepsilon)}{|1-\alpha^{2}(\varepsilon)r(\varepsilon)r^{\ast}(-\varepsilon)|^{2}}.\label{Rhedef}
\end{equation}
At the Fermi level $\varepsilon=0$ one has $\alpha=1$ and $rr^{\ast}=1-T$, hence $R_{he}=1$ irrespective of the transmission probability $T$ through the quantum dot. This is the Majorana resonance \cite{Law09}. Away from the Fermi level the resonance has (for $T\ll 1$) a Lorentzian decay $\propto[1+(\varepsilon/\Gamma)^{2}]^{-1}$, of width $\Gamma=T\delta_{\rm dot}/4\pi$ set by the average level spacing $\delta_{\rm dot}$ of the quantum dot.

The differential conductance $G=dI/dV$, at bias voltages $|V|<\Delta_{0}/e$ and in the zero-temperature limit, directly measures the probability \eqref{Rhedef}:
\begin{equation}
G/G_{0}=2+2R_{he}(eV),\;\;G_{0}=e^{2}/h.\label{Gresult}
\end{equation}
The two contributions to the conductance correspond to the two edges connecting the normal and superconducting contact: The edge containing the quantum dot contributes $2e^{2}/h\times R_{he}$, while the other edge remains unperturbed and contributes the full $2e^{2}/h$ --- for sufficiently small $B$ that the helical edge state remains gapless. 

The ensemble averaged conductance $\langle G\rangle$ has a peak value of $4G_{0}$ at $V=0$, above an off-resonant baseline $G_{\rm base}$ that we calculate as follows. We may assume $\delta_{\rm dot}\ll\Delta_{0}$, so we keep $\alpha=1$. We treat the off-resonant scattering amplitudes at $\pm\varepsilon$ as statistically independent random variables in the CUE, distributed according to Eq.\ \eqref{Pphidef}. Substitution of the parameterization \eqref{Sgeneric} into Eq.\ \eqref{Rhedef} gives, upon averaging,
\begin{align}
G_{\rm base}/G_{0}={}&2+2\int_{0}^{\pi/2}d\gamma_{+}\int_{0}^{\pi/2}d\gamma_{-}\,\sin 2\gamma_{+}\sin 2\gamma_{-}\nonumber\\
&\times\int_{0}^{2\pi}\frac{d\phi}{2\pi}\,\frac{\sin^{2}\gamma_{+}\sin^{2}\gamma_{-}}{|1-\cos\gamma_{+}\cos\gamma_{-}e^{i\phi}|^{2}}\nonumber\\
={}&\tfrac{2}{3}\pi^{2}-4\approx 2.58.\label{Gbase}
\end{align}
A similar calculation gives the \textit{triangular} line shape of $\langle G(V)\rangle$ as an average over the Lorentzian line shape of $G(V)$,
\begin{align}
\langle G(V)\rangle-G_{\rm base}&\propto\int_{0}^{1}dT\,[1+(4\pi eV/T\delta_{\rm dot})^{2}]^{-1}\nonumber\\
&=1-(4\pi eV/\delta_{\rm dot})\,{\rm arctan}\,(\delta_{\rm dot}/4\pi eV)\nonumber\\
&=1-2\pi^{2}e|V|/\delta_{\rm dot}+{\cal O}(V^{2}).\label{GVlineshape}
\end{align}

\begin{figure}[tb]
\centerline{\includegraphics[width=0.9\linewidth]{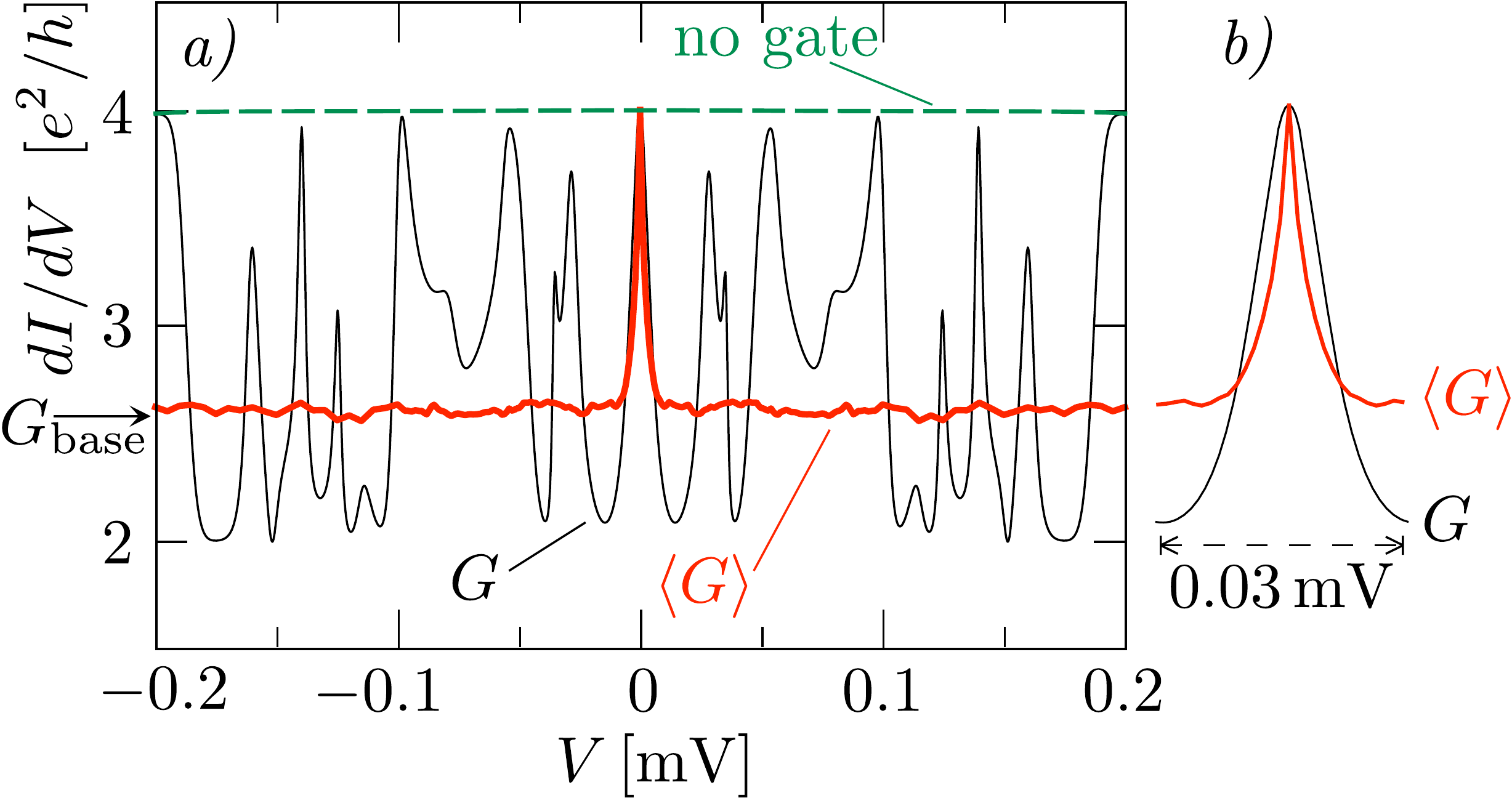}}
\caption{\textit{a)} Zero-temperature differential conductance $G=dI/dV$ as a function of bias voltage $V$, calculated numerically for the system of Fig.\ \ref{fig_setup}. The bottom of the conduction band in the gated region ($200\,{\rm nm}\times 200\,{\rm nm}$) is $E_{\rm c}=-1.5\,{\rm meV}$ below the Fermi level. The black curve is for a single disorder realization, the red curve is the disorder average. The calculated background conductance $G_{\rm base}$ from Eq.\ \eqref{Gbase} is indicated (arrow). For comparison, the conductance without the gate electrode is also shown (green dashed curve). The Majorana resonance is then fully absorbed in the background and invisible.
\textit{b)} Enlargement of the Majorana resonance from the left panel, to show the difference in line shape.
}
\label{fig_single}
\end{figure}

\begin{figure}[tb]
\centerline{\includegraphics[width=0.7\linewidth]{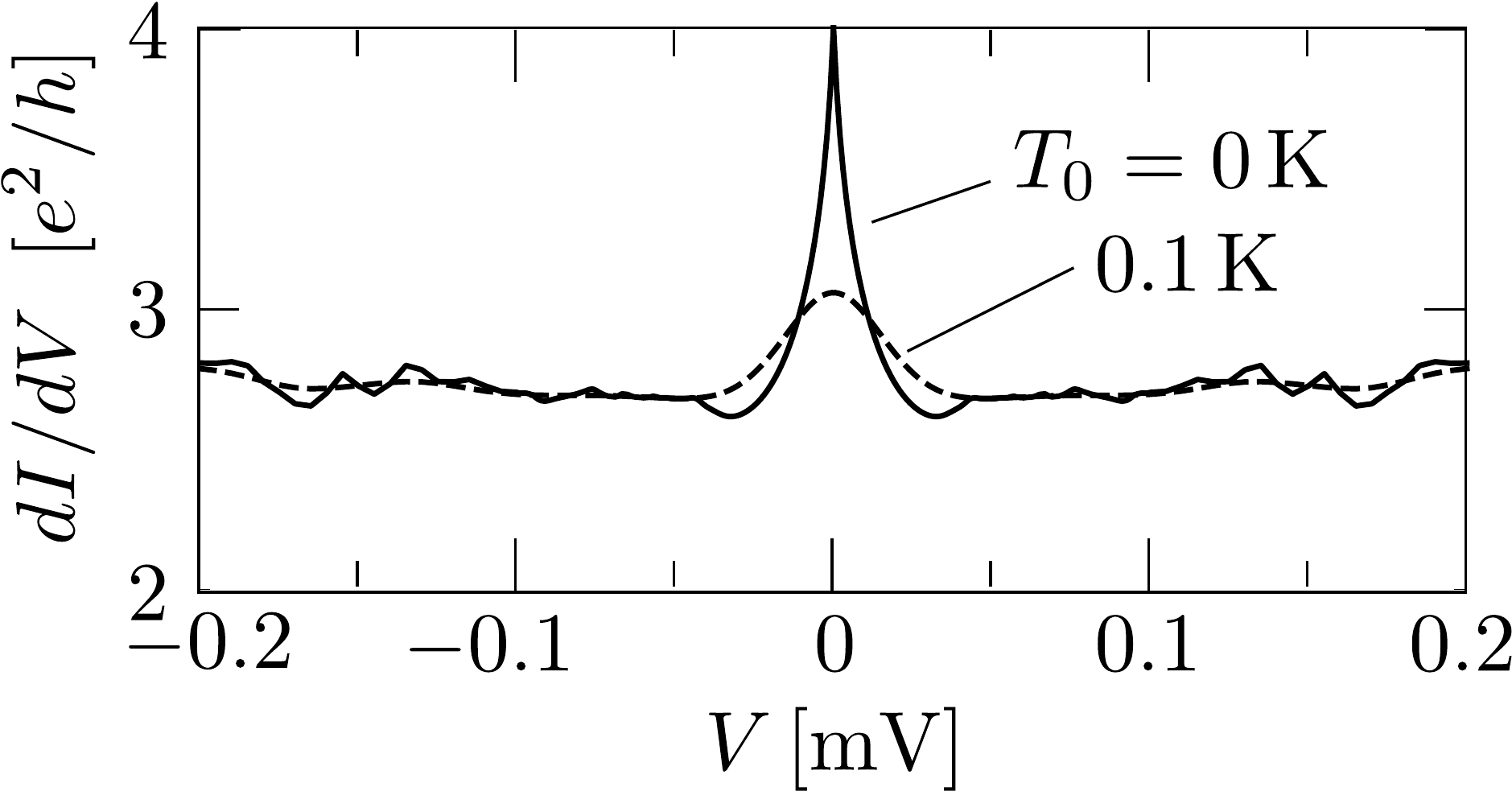}}
\caption{Differential conductance, averaged over gate voltages ($-4.5\,{\rm meV}\leq E_{\rm c}\leq-1.5\,{\rm meV}$) for a single disorder realization. All system parameters are the same as in Fig.\ \ref{fig_single}, except for the size of the gated region, which is $100\,{\rm nm}\times 100\,{\rm nm}$. The solid curve is at zero temperature and the dashed curve at a temperature of $0.1\,{\rm K}$.
}
\label{fig_gate}
\end{figure}

To test these analytical predictions, we have performed numerical simulations of a model Hamiltonian for an InAs/GaSb quantum well \cite{Liu08,Kne12,Liu08b,Kha11,note1}. Results are shown in Fig.\ \ref{fig_single} and fully confirm our expectations: Without the quantum dot the Majorana resonance remains hidden in the background conductance (dashed curve in Fig.\ \ref{fig_single}a), demonstrating that the $0.1\,{\rm T}$ applied field is weak enough to cause no appreciable backscattering of the helical edge states. We then create a $200\,{\rm nm}\times 200\,{\rm nm}$ quantum dot, as in Fig.\ \ref{fig_setup}, by applying a gate voltage. This suppresses the background conductance, revealing the Majorana resonance at $V=0$ (solid curves). Disorder averaging removes all resonances from Andreev levels at $V\neq 0$, so that the Majorana resonance stands out above the baseline conductance $G_{\rm base}$, in very good agreement with the calculated value \eqref{Gbase}. The triangular line shape of the average conductance is also confirmed by the simulations (Fig.\ \ref{fig_single}b).

The ensemble average in Fig.\ \ref{fig_single} is an average over disorder realizations. As is well known from quantum dot experiments \cite{Cha95,Kel96}, statistically equivalent ensembles may be generated for a fixed disorder potential by varying the gate voltage, which is more practical from an experimental point of view. In Fig.\ \ref{fig_gate} we show a computer simulation performed in this way. To reduce the sensitivity to thermal averaging, we took a smaller ($100\,{\rm nm}\times 100\,{\rm nm}$) quantum dot, keeping the magnetic field at $0.1\,{\rm T}$. The simulation shows that the Majorana resonance remains clearly visible above the background conductance at temperatures of 100~mK.

So much for the detection of Majorana zero-modes. In the final part of this paper, we take a longer term perspective and present a geometry that allows for the braiding of pairs of Majorana fermions, for the demonstration of the predicted non-Abelian statistics \cite{Rea00}. While the quantum spin Hall edge seems ideally suited for the detection of Majorana zero-modes, its one-dimensionality prevents the exchange of adjacent Majoranas. What is needed is a Y- or T-junction of superconductors to perform the ``three-point turn'' introduced by Alicea et al.\ \cite{Ali11} and implemented in a variety of braiding proposals for a network of nanowires \cite{Sau11,Hec12,Hal12,Hya13}.

\begin{figure*}[tb]
\centerline{\includegraphics[width=0.7\linewidth]{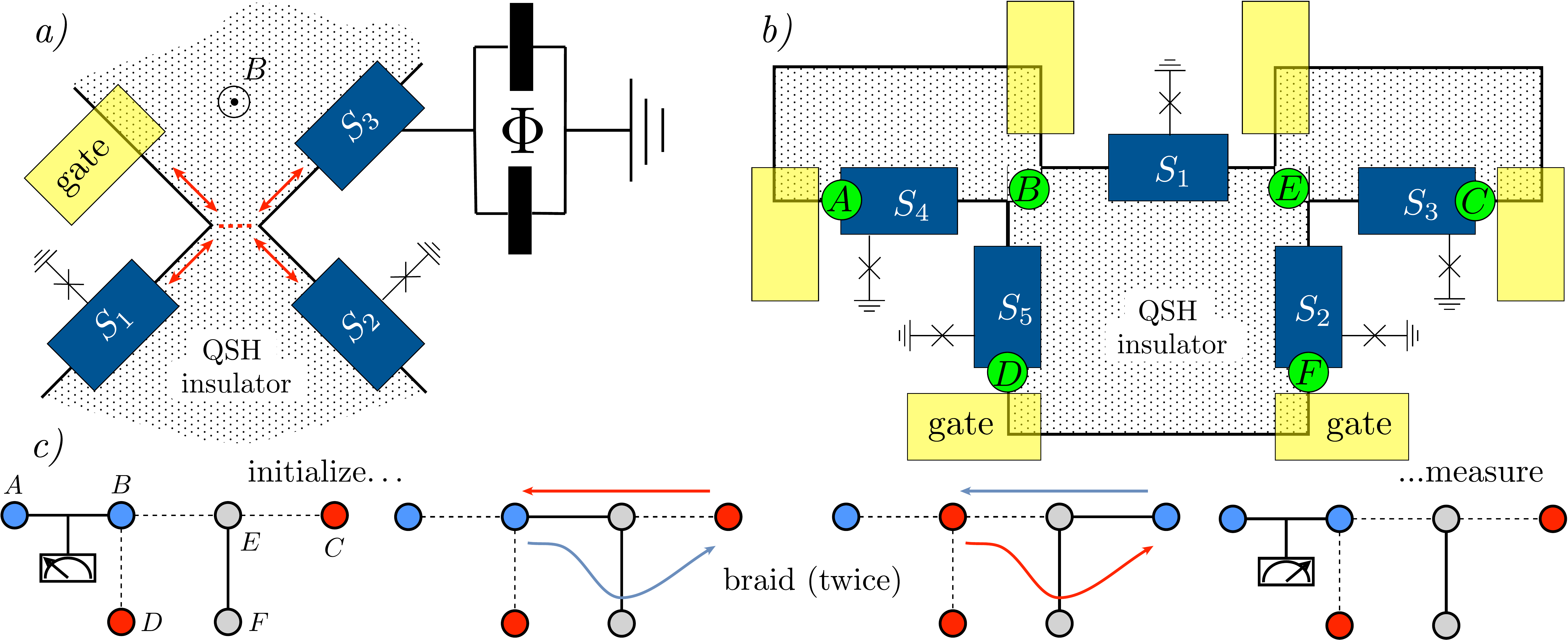}}
\caption{\textit{a)} Constriction in a quantum spin-Hall insulator, contacted along three edges by a superconducting island. If the fourth edge is blocked by a gate electrode, the constriction traps a Majorana fermion. Each superconducting island is connected to a superconducting ground by a split Josephson junction enclosing a magnetic flux, indicated schematically by a $\times$ symbol and shown expanded for one of the islands.
\textit{b)} Two constrictions in series form a $\pi$-shaped circuit that can be used to braid the Majorana fermions (green circles). The flux through each split Josephson junction controls the coupling of adjacent Majoranas.
\textit{c)} Schematic of the braiding operation in the $\pi$-circuit \cite{Hya13}. Coupled Majoranas are connected by a solid line, uncoupled Majoranas by a dashed line. 
}
\label{fig_setup_braiding}
\end{figure*}

In Fig.\ \ref{fig_setup_braiding}a we show how a constriction in the quantum spin Hall insulator can be used to achieve the same functionality as a crossing of nanowires. The constriction couples the helical edge states on opposite edges by tunneling, which is effective if it is narrower than the decay length of the edge states (100~nm or smaller). The coupling may be increased, if needed, by gating the constriction region into the conduction band. Three of the four edges leading into the constriction are gapped by a superconducting island. The fourth edge contains one of the quantum dots discussed earlier, tuned to a gate voltage interval of minimal transmission $T\ll 1$.

Let us check that the constriction traps a Majorana zero-mode. Helical edge states incident on the constriction from the three superconductors have reflection amplitudes that are contained in a $3\times 3$ reflection matrix $r(\varepsilon)$. Neglecting transmission throught the quantum dot, this is a unitary matrix. (A nonzero $T$ will give a finite width to the zero mode.) A bound state in the constriction at energy $\varepsilon$ is a solution of the determinantal equation \cite{Bee13}
\begin{equation}
{\rm Det}\,[1-\alpha^{2}\Lambda r(\varepsilon)\Lambda^{\ast}r^{\ast}(-\varepsilon)]=0,
\end{equation}
with $\Lambda={\rm diag}\,(e^{i\phi_1},e^{i\phi_2},e^{i\phi_3})$ a diagonal matrix containing the phase $\phi_{n}$ of the order parameter on the $n$-th superconductor. Since $\alpha(0)=1$, see Eq.\ \eqref{rAdef}, the condition for a zero mode is that the matrix $uu^{\ast}$, with $u\equiv\Lambda r(0)$, has an eigenvalue equal to $+1$. The eigenvalues of $uu^{\ast}$ come in complex conjugate pairs $e^{\pm i\psi}$. An unpaired eigenvalue at $-1$ is forbidden by ${\rm Det}\,uu^{\ast}=1$, but an unpaired eigenvalue at $+1$ is allowed and in fact necessary when the dimensionality of $u$ is odd --- as it is here.

In Fig.\ \ref{fig_setup_braiding}b,c we combine two constrictions in $\pi$-shaped circuit, to perform the braiding protocol of Ref.\ \cite{Hya13}. There are six Majorana fermions, one at each constriction and four more trapped by quantum dots along the quantum spin Hall edge. Adjacent Majorana operators $\gamma_{m},\gamma_{m'}$, for example $\gamma_B$ and $\gamma_E$, are coupled by the charging energy $E_{\rm C}$ of the intermediate superconducting island $S_n$ through the Hamiltonian \cite{Hec12}
\begin{equation}
H_{n}=iU_{n}\gamma_{m}\gamma_{m'},\;\;U_{n}\propto\exp\bigl[-\sqrt{8E_{\rm J}(\Phi_{n})/E_{\rm C}}\bigr],\label{Hndef}
\end{equation}
This coupling can be switched on and off by adjusting the Josephson energy $E_{\rm J}=E_{0}\cos(e\Phi/\hbar)$ of the superconducting island, via the magnetic flux $\Phi$ through a split Josephson junction that connects the island to a superconducting ground.

As worked out in Refs.\ \cite{Sau11,Hec12}, the alternating coupling and decoupling of adjacent Majoranas has the effect of exchanging them: One effectively moves the Majorana at $B$ through the T-junction towards $F$, followed by $C\mapsto B$ and finally $F\mapsto C$ completes the exchange of $B$ and $C$. If this exchange is repeated, one ends up with the original configuration of Majoranas, but in an orthogonal state: The fermion parity of $S_{4}$ has switched between even and odd. This signature of non-Abelian statistics can be measured as described in Ref.\ \cite{Hya13}, as a shift in the resonance frequency of a superconducting transmission line containing the circuit.

In conclusion, we have shown how the helical edge state in a quantum spin Hall insulator may be used as a single-channel, disorder-insensitive alternative to semiconductor nanowires, for the detection and braiding of Majorana fermions. For all we know, the experiments on InAs/GaSb quantum wells \cite{Kne12} may already have produced the predicted Majorana zero-modes \cite{Fu08}, but since the $4e^{2}/h$ conductance resonance is hidden in the $4e^{2}/h$ off-resonant background there is no way to tell. The quantum dot geometry proposed here lowers the average background to about $2.6\,e^{2}/h$, allowing for the emergence of the Majorana resonance. This seems to be an experiment that is fully within reach of existing devices, requiring only the addition of a nanostructured gate electrode and the application of a weak magnetic field. As a longer-term perspective, we have shown how a constriction in the quantum spin Hall insulator can reproduce the functionality of a nanowire T-junction, required for braiding and for the demonstration of non-Abelian statistics.

The numerical calculations were performed using the {\sc kwant} package developed by A. R. Akhmerov, C. W. Groth, X. Waintal, and M. Wimmer. We thank C. Liu for providing us with the model parameters of the InAs/GaSb quantum well. This work was supported by the Dutch Science Foundation NWO/FOM, by an ERC Advanced Investigator Grant, and by the China Scholarship Council.

\appendix

\section{Description of the numerical simulations}

Our numerical simulations are based on the four-band Hamiltonian of an InAs/GaSb quantum well \cite{Ber06,Liu08}, which in zero magnetic field takes the form
\begin{equation}
{\cal H}(\bm{k})=\begin{pmatrix}
H_{0}(\bm{k})&H_{1}(\bm{k})\\
-H_{1}^{\ast}(-\bm{k})&H_{0}^{\ast}(-\bm{k})
\end{pmatrix},\label{calHdef}
\end{equation}
as a function of wave vector $\bm{k}=(k_x,k_y)$ in the $x$-$y$ plane of the semiconductor layers. The block structure refers to the spin degree of freedom, while each block itself has a $2\times 2$ matrix structure that refers to the $(s,p)$ orbital degree of freedom. 

The diagonal block describes the hybridization of the \textit{s} and \textit{p} orbitals,
\begin{equation}
H_{0}=\begin{pmatrix}
U+b_{0}-b_{+}k^2&b_{3}k_+\\
b_{3}k_-&U-b_{0}-b_{-}k^2
\end{pmatrix},\label{H0def}
\end{equation}
with $k^2=k_x^2+k_y^2$, $k_{\pm}=k_x\pm ik_y$, and $b_{\pm}=b_{1}\pm b_{2}$. We have included an electrostatic potential $U$, to account for the effects of a gate electrode. The off-diagonal block describes the spin-orbit coupling by inversion asymmetry (Rashba and Dresselhaus effects),
\begin{equation}
H_{1}(\bm{k})=\begin{pmatrix}
c_{1}k_+ +ic_3 k_-&-c_0\\
c_0&c_{2}k_-
\end{pmatrix}.\label{H1def}
\end{equation}
to first order in $k$. The parameter values we used in our simulations, taken from Ref.\ \onlinecite{Liu08b}, are listed in Table I.

\begin{table}
\label{Tablequantumwell}
\begin{tabular}{l  | l  }
 \hline
$b_0 $ =            $-\hphantom{6}0.0078$ eV & $c_0$  =    $\hphantom{-}0.0002$   eV \\
$b_1$     =        $-\hphantom{6}5.8$    eV$\cdot$\AA$^2$ & $c_1$ =      $\hphantom{-}0.00066$ eV$\cdot$\AA \\
$b_2$        =     $-66.0$   eV$\cdot$\AA$^2$ & $c_2$  =    $\hphantom{-}0.0006$  eV$\cdot$\AA \\
$b_3$           =  $\hphantom{-6}0.37$    eV$\cdot$\AA & $c_3$      =   $-0.07$   eV$\cdot$\AA \\
  \hline
\end{tabular}
\caption{Parameters of the four-band Hamiltonian \eqref{calHdef}, representative for a heterostructure consisting of $10\,{\rm nm}$ InAs and $10\,{\rm nm}$ GaSb layers, sandwiched between AlSb barriers \cite{Liu08b}.
}
\end{table}

Time reversal symmetry is expressed by
\begin{equation}
{\cal H}(\bm{k})=\sigma_{y}{\cal H}^{\ast}(-\bm{k})\sigma_{y},\label{calTdef}
\end{equation}
where the Pauli matrix $\sigma_{y}$ acts on the spin blocks. A perpendicular magnetic field $\bm{B}=(0,0,B)$ breaks time reversal symmetry, via an orbital and a Zeeman effect. The orbital effect is accounted for by the substitution $\bm{k}\mapsto \bm{k}-(e/\hbar)\bm{A}$, with vector potential $\bm{A}=(0,Bx,0)$. The Zeeman energy $g\mu_{\rm B}B\sigma_{z}$ is a negligibly small effect in the weak magnetic fields $B\approx 0.1\,T$ considered here, so we do not include it.

The effect of a superconducting contact is introduced via the Bogoliubov-De Gennes Hamiltonian,
\begin{equation}
H_{\rm BdG}=\begin{pmatrix}
{\cal H}[\bm{k}-(e/\hbar)\bm{A}]&\Delta\\
\Delta^{\dagger}&-\sigma_{y}{\cal H}^{\ast}[-\bm{k}-(e/\hbar)\bm{A}]\sigma_{y}
\end{pmatrix}.\label{HBdGdef}
\end{equation}
The blocks of $H_{\rm BdG}$ refer to the $(e,h)$ electron-hole degree of freedom, coupled by the pair potential $\Delta$ induced by the superconducting contact. Electron-hole symmetry is expressed by
\begin{equation}
H_{\rm BdG}(\bm{k})=-(\sigma_{y}\otimes\tau_{y})H_{\rm BdG}^{\ast}(-\bm{k})(\sigma_{y}\otimes\tau_{y}),\label{ehsymm}
\end{equation}
where the Pauli matrix $\tau_{y}$ acts on the electron-hole blocks. A spin-singlet \textit{s}-wave proximity effect may still couple the \textit{s} and \textit{p} orbitals of the quantum well \cite{Kha11}, without breaking either electron-hole or time-reversal symmetry, but for simplicity here we take a scalar $\Delta$. 

For the numerical simulations we discretize the Hamiltonian \eqref{HBdGdef} on a square lattice (lattice constant $a=10\,{\rm nm}$) in the geometry shown to scale in Fig.\ \ref{fig_setup}. We set $\Delta=1\,{\rm meV}$ in the superconducting contact and zero elsewhere. The effect of the gate electrode is modeled by an offset $U_{\rm gate}$ of the electrostatic potential in the area under the gate. Disorder in the quantum well is modeled on the lattice by a random on-site potential $U_{\rm disorder}$, uniformly distributed in the interval $(-5\,{\rm meV},+5\,{\rm meV})$. The magnetic field in the quantum well and the normal-metal contact is fixed at $B=0.103\,{\rm T}$, corresponding to one flux quantum $h/e$ through an area of size $200\,{\rm nm}\times 200\,{\rm nm}$. Inside the superconducting contact we set $B=0$, ignoring the penetration of flux in magnetic vortices.

At excitation energies $|\varepsilon|<\Delta$ the electrons and holes incident on the superconductor are fully reflected. The reflection amplitudes are contained in an $8\times 8$ unitary reflection matrix. The $4\times 4$ Andreev reflection block $r_{he}(\varepsilon)$ gives the differential conductance at zero temperature,
\begin{equation}
G(V,0)=\frac{2e^{2}}{h}{\rm Tr}\,r_{he}^{\vphantom{\dagger}}(eV)r_{he}^{\dagger}(eV).\label{GcalRdef}
\end{equation}
The corresponding result at finite temperature $T_{0}$ follows upon integration,
\begin{equation}
\begin{split}
&G(V,T_{0})=-\int_{-\infty}^{\infty}d\varepsilon\,G(\varepsilon/e,0)\frac{\partial f}{\partial\varepsilon},\\
&f(\varepsilon,V,T_{0})=\frac{1}{1+\exp[(\varepsilon-eV)/k_{\rm B}T_{0}]}.
\end{split}
\label{GVfiniteT}
\end{equation}

\end{document}